\documentclass[aps,amsfonts,pra,twocolumn,showpacs]{revtex4-1}
\usepackage{epsfig,amsmath,amssymb,bm,epsf,graphicx,psfrag,float}
\usepackage[all]{xy}
\usepackage{color}

\def\ket#1{\vert#1\rangle}
\def\ketbra#1{\vert#1\rangle\langle#1\vert}

\def\Longarrow{\protect\@lra}
\def\@lra{\relbar\joinrel\relbar\joinrel\relbar\joinrel%
          \relbar\joinrel\rightarrow}

\newcommand{\bc}{\begin{center}}
\newcommand{\ec}{\end{center}}
\newcommand{\be}{\begin{equation}}
\newcommand{\ee}{\end{equation}}
\newcommand{\bea}{\begin{eqnarray}}
\newcommand{\eea}{\end{eqnarray}}

\newcommand{\ncd}{\newcommand}
\ncd{\QCcns}{$QC_{\cal{C}}$}
\ncd{\QCc}{$QC_{\cal{C}}\;$}

\definecolor{libl}{cmyk}{0.2,0.1,0,0}

\begin{document}

\title{Quantum computational universality of Affleck-Kennedy-Lieb-Tasaki states
beyond the honeycomb lattice}

\author{Tzu-Chieh Wei}
\affiliation{C. N. Yang Institute for Theoretical Physics and the
Department of Physics and Astronomy, State University of New York at
Stony Brook, Stony Brook, NY 11794-3840, USA}

\date{\today}

\begin{abstract}
Universal quantum computation can be achieved by simply performing
single-spin measurements on a highly entangled resource state, such
as cluster states. The family of Affleck-Kennedy-Lieb-Tasaki (AKLT)
states has recently been explored; for example,  the spin-1 AKLT
chain can be used to simulate single-qubit gate operations on a
single qubit, and the spin-3/2 two-dimensional AKLT state on the
honeycomb lattice can be used as a universal resource. However, it
is unclear whether such universality is a coincidence for the
specific state or a shared feature in all two-dimensional AKLT
states. Here we consider the family of spin-3/2 AKLT states on
various trivalent Archimedean lattices and show that in addition to
the honeycomb lattice, the spin-3/2 AKLT states on the square
octagon $(4,8^2)$ and the `cross' $(4,6,12)$ lattices are also
universal resource, whereas the AKLT state on the `star' $(3,12^2)$
lattice is likely not due to geometric frustration.
\end{abstract}
 \maketitle

 \section{ Introduction} Measurement-based quantum computation (MBQC) is an alternative
 approach to the standard circuit model in realizing quantum computation, which promises exponential
speedup over classical computation~\cite{NielsenChuang00}.  In the
former, local measurement alone achieves the same power of
computation, provided a prior sufficiently entangled state is
given~\cite{Oneway,Oneway2,RaussendorfWei12}. One of the challenges
in MBQC is to identify these entangled states, namely, the {\it
universal resource states\/}, that enable the success of driving
universal quantum computation by local measurement. If a state
possesses too little entanglement, naturally, it cannot provide
sufficient quantum correlation to drive universal quantum
computation~\cite{VandenNestDurVidalBriegel07,VandenNest}. On the
other hand, if a state possesses too much entanglement, the
measurement outcome cannot provide any advantage over classical
random guessing~\cite{Gross1}.  Universal resource states are thus
found to be very rare~\cite{Gross1}.

 The first discovered resource state is the cluster
state on the 2D square lattice~\cite{Oneway,Cluster}. However,
cluster states do not arise as unique ground states of two-body
interacting Hamiltonians~\cite{Nielsen}. This is a disadvantage from
the viewpoint of creating universal resource states by cooling. By
 reverse engineering of finding
suitable two-body parent Hamiltonians, these universal resource
states can arise as unique ground
states~\cite{Gross,Verstraete,Chen}.
 Recently it was discovered that the
one-dimensional spin-1 Affleck-Kennedy-Lieb-Tasaki (AKLT)
state~\cite{AKLT,AKLT2} can be used to implement arbitrary one-qubit
rotations on a single qubit~\cite{Gross,Brennen}. The AKLT state was
originally constructed to support Haldane's conjecture regarding the
spectral gap of integer-spin Heisenberg
chains~\cite{Haldane,Haldane2}. More generally, AKLT states can be
defined on vertices of any graph or lattice. With suitable boundary
conditions they are the unique ground states of
Hamiltonians~\cite{KLT}.  The AKLT Hamiltonians have
nearest-neighbor two-body interactions respecting the rotational
symmetry of spins and the spatial symmetry of the underlying
lattice.  The usefulness of AKLT states opens new avenues for
experimental realization~\cite{Resch} and has instilled novel
concepts in MBQC, such as the renormalization group~\cite{Bartlett},
the holographic principle~\cite{Miyake} and the symmetry-protected
topological orders~\cite{ElseSchwarzBartlettDoherty}.

\begin{table}[t!]
  \begin{tabular}{|c|c|c|c|}
  \hline
  & honeycomb & square-octagon & cross \\
  \hline
  vertex degree & $3.520(1)$ & $3.065(1)$ & $2.980(1)$ \\
  \hline
  ave domain size & 2.016(1) & 2.025(2) & 2.029(2)\\
  \hline
  no. of vertices & $0.495(1) N$ & $0.488(1)N$ & $0.485(1)N$ \\
  \hline
  no. of edges & $0.871(1) N$ & $0.748(1) N$ & $0.723(1) N$\\
  \hline
  $p_{\rm th}^{\rm [site]}$ &  0.67(1) & 0.74(1) & 0.79(1) \\
  \hline
  $p_{\rm th}^{\rm [bond]}$ &  0.57(1) & 0.65(1) & 0.71(1) \\
  \hline
  \end{tabular}
 \caption{\label{tbl:graph} Properties of typical graphs resulting
 from the POVM on AKLT states residing in three different trivalent lattices. The numbers are extracted to the asymptotic limit (i.e., large site number $N$). The results of the honeycomb lattice~\cite{WeiAffleckRaussendorf11} are included for comparison. AKLT states on all three lattices are universal.
 In contrast, the AKLT state on the star lattice is likely not a universal resource.}
\end{table}
However, to achieve universal quantum computation within the
measured-based architecture a two-dimensional structure or higher is
needed. This is because a single spin chain can only be used to
simulate the single-qubit gates on one qubit and thus multiple
chains are needed for quantum computation. Moreover, entanglement
between qubits on different chains is needed to implement controlled
gates. Can any 2D AKLT state be used for universal quantum
computation? Cai et al. utilized many 1D AKLT quasichains to form a
2D structure~\cite{Cai10} by merging pairs of spin-1/2 particles on
neighboring quasichains. They showed that the resulting state, even
though it is no longer an AKLT state, is a universal resource state.
Its universality was also understood in terms of both the ability to
generate a one-dimensional cluster state from an AKLT qusaichain and
the additional ability to further form a two-dimensional cluster
state by measuring the merged spins between neighboring
quasichains~\cite{WeiRaussendorfKwek11}.

\begin{figure}
   \includegraphics[width=8cm]{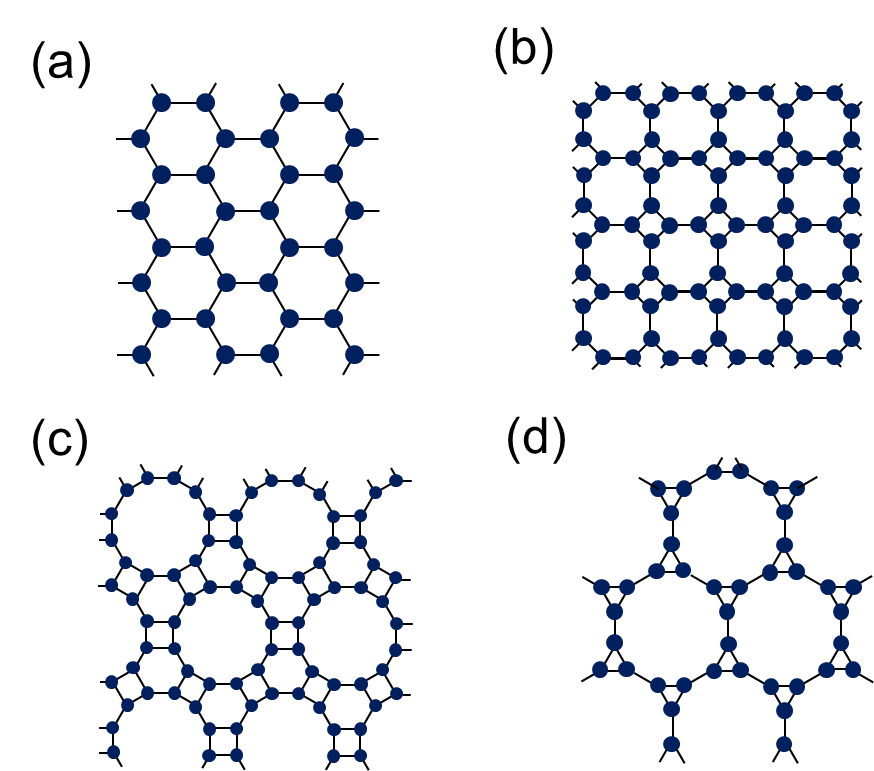}
  \caption{\label{fig:lattices} (color online) The four trivalent Archimedean lattices. (a) the honeycomb $(6^3)$; (b) the square octagon $(4,8^2)$;
  (c) the cross $(4,6,12)$; and (d) the star $(3,12^2)$. }
\end{figure}

It was shown, by the author and his
collaborators~\cite{WeiAffleckRaussendorf11} and independently by
Miayke~\cite{Miyake11}, that the intrinsic 2D AKLT state on the
honeycomb lattice indeed provides a universal resource for MBQC. The
notion of the computational backbone for universal quantum
computation was introduced by Miyake in Ref.~\cite{Miyake11} and his
argument that such backbone exists in the honeycomb lattice seems
intuitive. The key point of his argument is that the backbone
consists mainly of the
 sites that have different POVM outcomes with their neighbors
as well as the edges that link these sites~\cite{Miyake11}. For a
pair of neighboring sites, the six outcomes: $xy$, $xz$, $yx$, $yz$,
$zx$, $zy$ are desirable whereas the remaining three $xx$, $yy$,
$zz$ do not contribute to the backbone. This corresponds
approximately  to the probability that an edge occupied by the
backbone being $p_{\rm occ}=2/3$. Because the bond percolation
threshold for the honeycomb lattice is $p_{th}^{\rm (bond)}\approx
0.65 < p_{\rm occ}$, the computational backbone spanning across the
lattice exists and thus the AKLT state is a universal resource.
However, as we shall see below that the argument cannot extend to
other trivalent lattices with $p_{th}> 2/3$.

So far the 2D AKLT state on the honeycomb lattice is the only state
in the AKLT family that has been shown to be a universal resource.
It is unclear whether its computational universality is a
coincidence or a more general feature in the family of
two-dimensional AKLT states. From the viewpoint of magnetic
ordering, in three dimensions, AKLT states can be N\'eel ordered on
many lattices such as the cubic lattice, but they can also be
disordered on lattices such as diamond and pyrochlore~\cite{Param}.
But all 2D AKLT states are found to be disordered. N\'eel ordered
states are believed not to possess sufficient quantum correlation
for the MBQC. Is being disordered in 2D or higher sufficient for the
AKLT states to be universal for MBQC? In contrast for cluster
states, it has been shown that all 2D cluster states (i.e., graph
states on all 2D regular lattices; see also
Sec.~\ref{sec:2Dcluster}), including those on the square,
triangular, honeycomb, and Kagome lattices, are all
universal~\cite{Universal}.

 Here, we focus ourselves
on the family of spin-3/2 AKLT states but on various trivalent
Archimedean lattices, which include the honeycomb $(6^3)$, the
square octagon $(4,8^2)$, the `cross' $(4,6,12)$ and the `star'
$(3,12^2)$ lattices; see Fig.~\ref{fig:lattices}. All latter three
lattices have bond percolation thresholds higher than $2/3$, in
contrast to the honeycomb lattice, which has the threshold
approximately $0.65$. Does this mean that AKLT states on these
trivalent lattices but the honeycomb are {\it not\/} universal,
i.e., cannot be used to implement universal MBQC? Surprisingly, as
we shall find below, in addition to the honeycomb lattice, the
spin-3/2 AKLT states on the square octagon and the cross lattices
are also universal resources, whereas the spin-3/2 AKLT state on the
star $(3,12^2)$ lattice is likely not due to geometric frustration;
see Table~\ref{tbl:graph} for a summary. Along the way, we have also
provided an efficient method for the Metropolis update, improving on
the accessible system sizes that previous works could
achieve~\cite{WeiAffleckRaussendorf11,WeiAffleckRaussendorf12,DarmawanBrennenBartlett}.

The remaining of the paper is organized as follows. In
Sec.~\ref{sec:AKLTcluster} we define the AKLT, graph and cluster
states. In Sec.~\ref{sec:AKLT2graph} we describe the procedure to
convert an AKLT state to a graph state by POVM. If the resulting
graph state lies in the supercritical phase, then it can be used for
universal quantum computation. In Sec.~\ref{sec:numerics} we
describe an efficient Metropolis update method and perform numerical
simulations for AKLT states on various trivalent Archimedean
lattices. Unlike the other lattices (honeycomb, square octagon and
cross), the star lattice possesses geometric frustration and the
associated AKLT state is argued to fail to provide a universal
resource. In Sec.~\ref{sec:conclude} we make some concluding
remarks.
\begin{figure}
   \includegraphics[width=8cm]{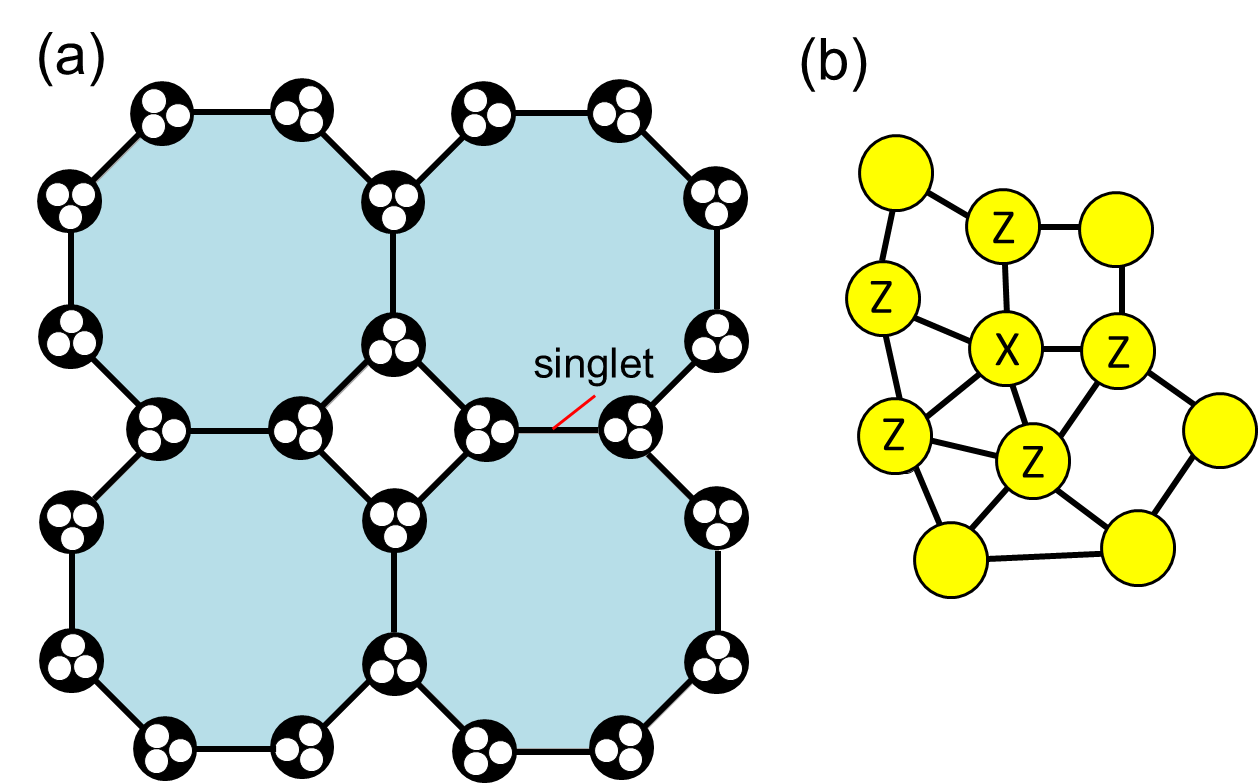}
  \caption{\label{fig:states}
  (color online) Illustrations of the AKLT state on a square-octagon lattice and a  graph state on a graph. (a) AKLT
  state. Each site is spin-3/2 and can be regarded as three virtual
  spin-1/2's in their symmetric subspace.
  Two virtual qubits connected by an edge form a spin singlet state.
  (b) A graph state. One qubit  resides at each vertex and one stabilizer generator of
   the form $X\otimes Z\cdots Z$ is shown (with $X\equiv\sigma_x$ at a vertex and $Z\equiv\sigma_z$ on
  neighboring vertices). Cluster states
  are graph states defined on regular lattices.}
\end{figure}

\section{AKLT and cluster states}
\label{sec:AKLTcluster} Let us now define the spin-3/2 AKLT states,
the graph and cluster states. The AKLT state~\cite{AKLT} on the
trivalent Archimedean lattices ${\cal{L}}$ (see
Fig.~\ref{fig:lattices})
 has
one spin-3/2 per site of lattice ${\cal{L}}$. The state space of
each spin 3/2 can be viewed as the symmetric subspace of three
virtual spin-1/2's, i.e., qubits. In terms of these virtual qubits,
the AKLT state on ${\cal{L}}$ is (see Fig.~\ref{fig:states}a)
\begin{equation}
  \label{AKLT2}
  |\Phi_{\rm AKLT}\rangle\equiv\bigotimes_{v \in V({\cal{L}})}P_{S,v}
\bigotimes_{e \in E({\cal{L}})} |\phi\rangle_e,
\end{equation}
where $V({\cal L})$ and $E({\cal L})$ to denote the set of vertices
and edges of ${\cal L}$, respectively. $P_{S,v}$ is the projection
onto the symmetric (equivalently, spin 3/2) subspace at site $v$ of
${\cal{L}}$~\cite{WeiAffleckRaussendorf11,WeiAffleckRaussendorf12}.
For an edge $e=(v,w)$, $|\phi\rangle_{e}$ denotes a singlet state,
with one virtual spin 1/2 at vertex $v$ and the other at $w$.
 On the other hand, a graph
state $\ket{G}$ is a stabilizer state~\cite{Hein} with one qubit per
vertex of the graph $G$ (see Fig.~\ref{fig:states}b) and  is the
unique eigenstate of a set of commuting operators~\cite{Cluster},
usually called the stabilizer generators~\cite{Stabilizer},
\begin{equation}
\label{eqn:stabilizerGen} \sigma_x^{[v]}\bigotimes_{u\in {\rm
nb}(v)} \sigma_z^{[u]}\,\, \ket{{G}}=\ket{{G}}, \ \forall v\in
V({G}),
\end{equation}
where ${\rm nb}(v)$ denotes the neighbors of vertex $v$, and
$\sigma_x$, $\sigma_y$ and $\sigma_z$ are the
 Pauli matrices. The original cluster state was defined as the graph state on the square lattice,
 but recent trend is to refer to any graph state
with the underlying graph being a regular lattice as a cluster
state. We shall adopt this broader definition of a cluster state.

\subsection{Remarks on universality of 2D cluster states and other graph states}
\label{sec:2Dcluster}
 It was shown by Van den Nest et al.~\cite{Universal} that the
graph states on the
 triangular, honeycomb, and Kagome
lattices can all be converted to graph states on the square lattice
via local measurement, hence proving that these cluster states also
provide the same quantum computational power as the square-lattice
cluster state.  Their approach can be applied to show that the
cluster states on the square-octagon, cross and star lattices are
also universal, the proof of which is given in Appendix A. Browne et
al.~\cite{Browne} considered faulty cluster states on the square
lattice, i.e., graph states whose graphs are obtained from square
lattice by site deletion and are thus irregular. Such irregular
graph states (from site deletion) are universal if the site
occupation probability is below the site percolation threshold of
the square lattice. Their treatment was subsequently generalized to
general 2D random graphs, and naturally the universality of the
corresponding graph states arises if the graphs are in the
supercritical phase of percolation~\cite{WeiAffleckRaussendorf12}.
On the other hand, even though AKLT states can be defined on any
graph, their fate of universality is much less understood. As
detailed below, our proof of universality for the trivalent AKLT
states relies on reduction of them to graph/cluster states via a
particular local generalized measurement~(\ref{POVM2}). If the
reduction from AKLT states to universal graph states can be shown,
then the universality of the AKLT states is thus established.
However, to prove that a particular AKLT state is not universal, one
may need to show that for all possible local measurement, such a
reduction to universal graph states is not possible. Other possible
approaches for proving non-universality include the scaling of
entanglement~\cite{Universal} and efficient classical
simulation~\cite{VandenNestDurVidalBriegel07}. These latter methods
seems unlikely useful for proving non-universality for 2D AKLT
states, as they possess entanglement that satisfies an area law and
the tensor network description for all trivalent AKLT states gives
essentially the same local tensors.
\begin{figure*}
   \begin{tabular}{ll}
    (a) & (b)\\
    \hspace*{0.3cm}
  \includegraphics[width=8cm]{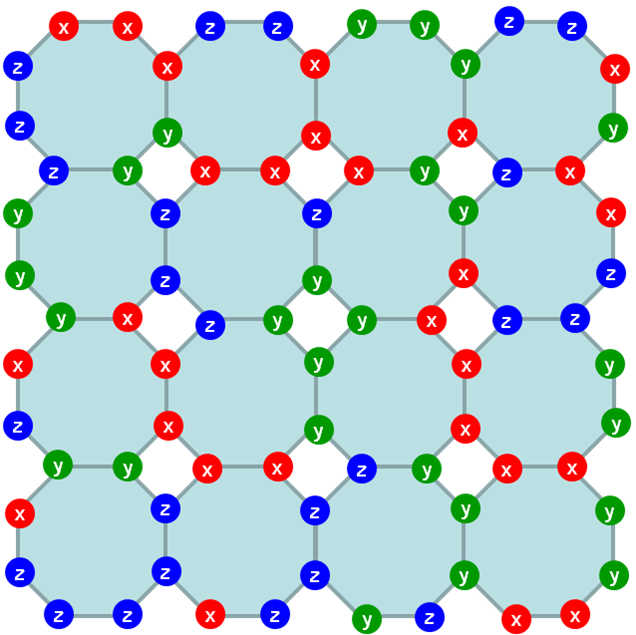}&  \hspace*{0.3cm}
  \includegraphics[width=8cm]{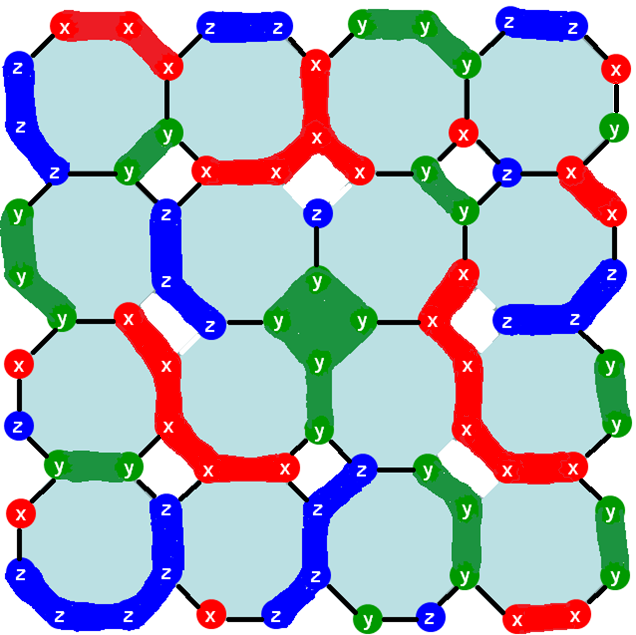}
   \end{tabular}
  \caption{\label{fig:sqoc} (color online) Illustration of POVM outcomes and construction of the graph for the resulting graph state on a square-octagon lattice. (a) POVM outcomes: x (red), y (green) and z (blue). (b) The graph after formation of domains and edge modulo-2 operation.
 }
\end{figure*}

\section{From AKLT  to graph states}
\label{sec:AKLT2graph} 
\begin{figure*}
   \begin{tabular}{ll}
    (a) & (b)\\
    \hspace*{0.3cm}
  \includegraphics[width=8cm]{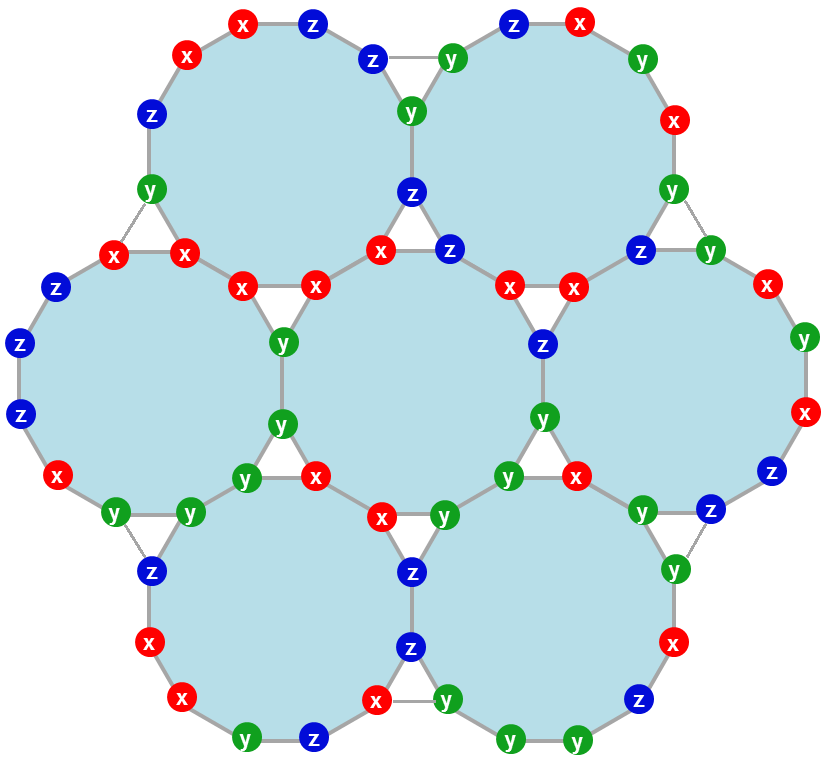}& \hspace*{0.3cm}
   \includegraphics[width=8cm]{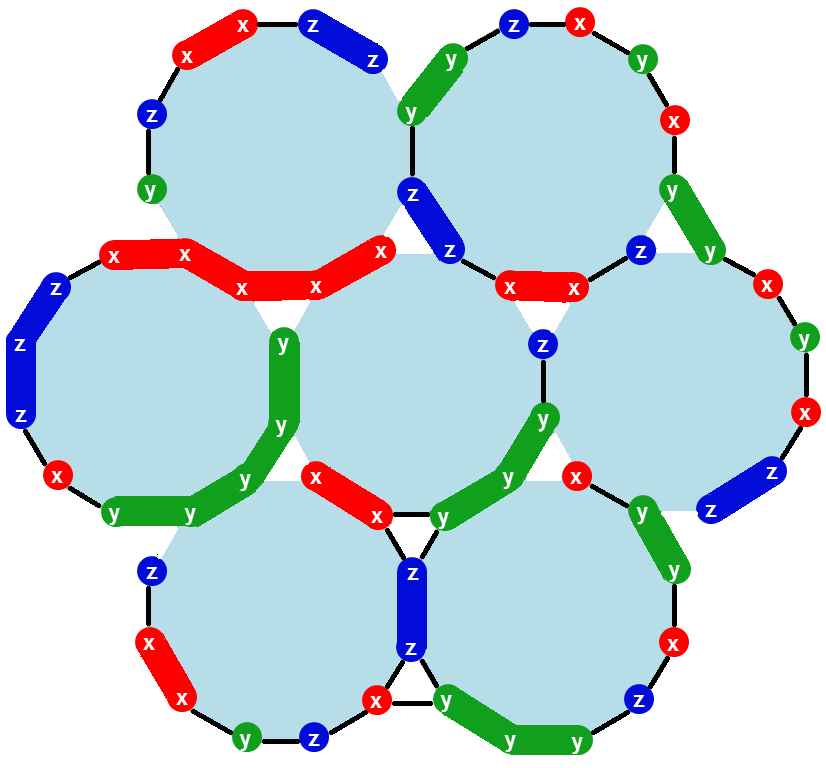}
   \end{tabular}
  \caption{\label{fig:star}(color online) Illustration of POVM outcomes and construction of the graph for the resulting graph state on a star lattice. (a) POVM outcomes: x (red), y (green) and z (blue). (b) The graph after formation of domains and edge modulo-2 operation.
One can also see the effect of the geometric frustration, which
causes two edges in some triangles to be removed. }
\end{figure*}

The AKLT state on the honeycomb lattice was
shown, via two different approaches, in
Ref.~\cite{WeiAffleckRaussendorf11} and Ref.~\cite{Miyake11},
respectively, to be a universal resource. In the
former~\cite{WeiAffleckRaussendorf11} a generalized measurement or
positive-operator-value-measure measurement (POVM), see below, on
all spins gives rises to a random planar graph state
$|G({\cal{A}})\rangle$, whose graph $G({\cal{A}})$  possesses a
macroscopic number of vertices and connections and from which a 2D
cluster state can be carved out by local measurement. Here, ${\cal
A}$ denotes the set measurement outcomes on all sites; see
Figs.~\ref{fig:sqoc} and~\ref{fig:star} for example. In the
latter~\cite{Miyake11} the same POVM was used but universal gates
were constructed and a computational backbone on the honeycomb
lattice was argued to exist. Its existence relies on the bond
percolation threshold of the honeycomb lattice being smaller than
$2/3$, as explained earlier. For all the lattices that we shall
examine, their bond percolation thresholds are all higher than $2/3$
(approximately, $0.6768$ for the square-octagon, $0.6737$ for the
cross, and $0.7404$ for the star, respectively). This suggests that
either the AKLT states on these lattices may not be universal for
MBQC or the argument used in Ref.~\cite{Miyake11} simply cannot be
extended to these cases. Therefore,
 to determine whether they can be universal,
 we shall follow the approach employed in Ref.~\cite{WeiAffleckRaussendorf11},
 where the argument
that, after the POVM, the AKLT state is mapped to a random planar
graph state $|G({\cal{A}})\rangle$ applies more generally and,
specifically, to other trivalent AKLT states. The MBQC on cluster
and graph states rely on the ability to simulate multiple-qubit
evolution along quantum wires and interaction between qubits on
neighboring wires. The computational universality of typical graph
states $|G({\cal{A}})\rangle$ hence is dependent on the connectivity
of $G({\cal{A}})$, and thus the problem whether these AKLT states
are universal resource can be resolved by examining the properties
of the typical graphs resulting from the POVM.

 Let us elaborate on one of the key steps in
 the construction, i.e., the POVM. The approach we shall use is to convert
 locally four-level spin-3/2 particles to two-level qubits, and this requires projection
 or,
 more generally, a local generalized
measurement~\cite{NielsenChuang00} (or POVM), on every site $v$ on
${\cal{L}}$. The POVM consists of three rank-two elements
\begin{subequations}
\label{POVM2}
  \begin{eqnarray}
\!\!\!\!
{F}_{v,z}\!\!\!\!\!\!\!\!&&=\sqrt{\frac{2}{3}}(\ketbra{000}+\ketbra{111}) \\
\!\!\!\!\!\!\!\!\!\!&=&\sqrt{\frac{2}{3}}\Big(\ketbra{S_z=\frac{3}{2}}+\ketbra{S_z=-\frac{3}{2}}\Big)\nonumber
 \\
\!\!\!\!{F}_{v,x}\!\!\!\!\!\!\!\!&&=\sqrt{\frac{2}{3}}(\ketbra{+++}+\ketbra{---})\\
\!\!\!\!\!\!\!\!\!\!&=&\sqrt{\frac{2}{3}}\Big(\ketbra{S_x=\frac{3}{2}}+\ketbra{S_x=-\frac{3}{2}}\Big)\nonumber
 \\
\!\!\!\!{F}_{v,y}\!\!\!\!\!\!\!\!&&=\sqrt{\frac{2}{3}}(\ketbra{i,i,i}+\ketbra{-\!i,-\!i,-\!i})\\
\!\!\!\!\!\!\!\!\!\!&=&\sqrt{\frac{2}{3}}\Big(\ketbra{S_y=\frac{3}{2}}+\ketbra{S_y=-\frac{3}{2}}\Big)\nonumber,
\end{eqnarray}
\end{subequations}
where $|0/1\rangle$, $\ket{\pm}\equiv(\ket{0}\pm\ket{1})/\sqrt{2}$
and $\ket{\pm i}\equiv (\ket{0}\pm i\ket{1})/\sqrt{2}$ are
eigenstates of the Pauli operators $\sigma_z$, $\sigma_x$ and
$\sigma_y$, respectively. As indicated in the second line of each of
the above three equations, physically, $F_{v,a}$ is proportional to
a projector onto the two-dimensional subspace spanned by the
$S_a=\pm 3/2$ states. Its action on the rotationally invariant AKLT
state gives a locally preferred quantization axis $a$. The above
POVM elements obey the relation $\sum_{\nu \in
\{x,y,z\}}F^\dagger_{v,\nu} F_{v,\nu} = P_{S,v}$ in terms of three
qubits, i.e., the sum projects onto the symmetric subspace, as
required, or equivalently equals identity in terms of spin-3/2
states. Such POVM can, in principle, be implemented by coupling the
local spin (whose initial state denoted by $|\psi\rangle_v$) to a
meter (whose initial state denoted by $|0\rangle_m$) so that the
spin and meter involve unitarily:
\begin{equation}
\label{eqn:U} U: \ |\psi\rangle_v\otimes |0\rangle_m \rightarrow
\sum_{a=x,y,z} F_{v,a}|\psi\rangle_v \otimes |a\rangle_m,
\end{equation}
where $|x\rangle_m$, $|y\rangle_m$, and $|z\rangle_m$ are three
orthogonal meter states. Upon `reading' the meter state to be
$|a\rangle_m$, the spin state `collapses' to
$F_{v,a}|\psi\rangle_v$. We note that the unitary transformation $U$
in Eq.~(\ref{eqn:U}) can always be found and is not
unique~\cite{NielsenChuang00}.

In contrast to usual projection operators where there are some
probability of getting undesired outcomes (i.e., lying in the
orthogonal subspaces to the desired outcomes), here any outcome of
the POVM is a successful mapping from four levels to two levels. The
outcome $a_v$ of the POVM at any site $v$ is random, which can be
$x$, $y$ or $z$, but it is correlated with the outcomes at other
sites due to quantum correlations in the AKLT state~\cite{AKLT}. As
shown in
Refs.~\cite{WeiAffleckRaussendorf11,WeiAffleckRaussendorf12}, the
resulting quantum state, dependent on the random POVM outcomes
${\cal{A}}=\{a_v, v\in V({\cal{L}})\}$,
\begin{equation}
  \label{CA}
  |\Psi({\cal{A}})\rangle = \bigotimes_{v \in V({\cal{L}})} \!\! F_{v,a_v}\, |\Phi_{\rm AKLT}
\rangle
\end{equation}
is equivalent to an encoded graph state ${|G({\cal{A}})\rangle}$.

One important consequence of the POVM is that a logical qubit, which
sits on a
 {\it domain\/} (a vertex of $G({\cal{A}})$), can be composed of multiple spins (i.e. those
spins whose sites are connected and whose POVM outcomes are the
same). Such encoding originates from the antiferromagnetic property
inherent in AKLT states: neighboring spin-3/2 particles must not
have the same $S_a=3/2$ (or -3/2) configuration~\cite{AKLT}. Hence,
after the projection onto $S_a=\pm3/2$ subspace by the POVM, the
configurations for all sites inside a domain can only be
$\ket{3/2,-3/2,3/2,-3/2,\dots}$ or $\ket{-3/2,3/2,-3/2,3/2,\dots}$,
and these form the basis or the encoding of a {\it single\/} qubit.
This corresponds to that a domain is formed by contracting all edges
that connect sites with the same POVM outcome. Equivalently, those
edges that are contracted become {\it internal\/} edges of a domain.
We note that even though a logical qubit is encoded by several
physical spins, the quantum information it carries can be
concentrated to any one of the spins, by measuring other spins in
the basis $\{\ket{3/2}\pm \ket{-3/2}\}$, still requiring only local
measurements.

The graph for the corresponding graph state is formed by domains as
vertices. What about the (external) edges?
 When multiple sites are merged into a domain, two domains can share multiple edges. It turns out that
 in accordance with the fact that Pauli $\sigma_z$ squares to identity (i.e. $\sigma_z^2=I$), the edges between two domains should
 be regarded in a modulo two fashion (i.e. edge modulo-2 operation):  one should delete
all edges of even multiplicity and convert all edges of odd
multiplicity into one single edge. See Figs.~\ref{fig:sqoc}
and~\ref{fig:star} for illustration of domain formation and edge
modulo-2 operation.

\begin{figure}
   \includegraphics[width=8cm]{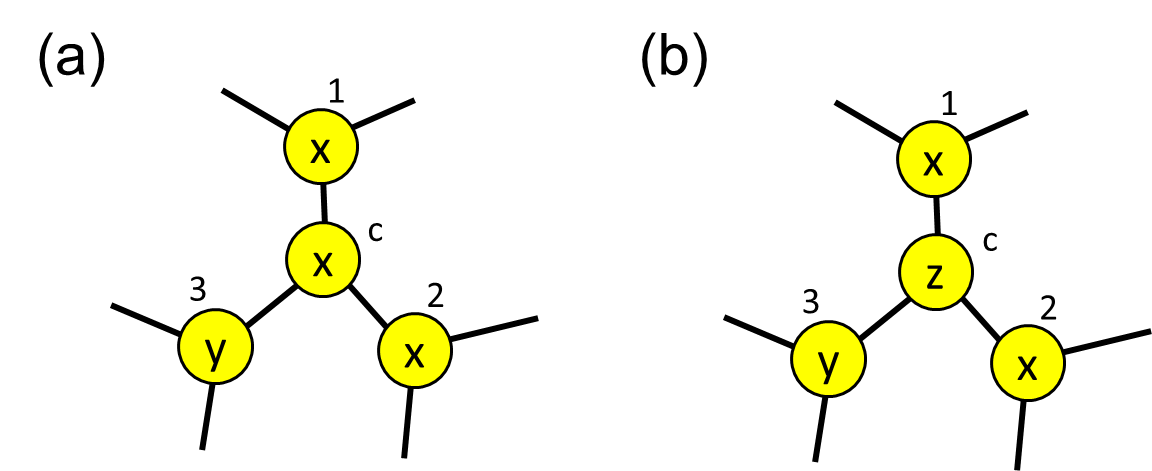}
  \caption{\label{fig:wolff}
 (color online) Illustrations of the counting of $e_I$ and $v_c$ (see text).  The POVM
 types are shown in the circles: (a) before the type at $c$ was flipped;
 (b) after it was flipped.
  In (a), $e_I=2$, arising from edges $(c,1)$ and $(c,2)$. In (b) the type on $c$ is
  flipped from $x$ to $z$ and it is easy to see that $e_I=0$. In (a), there are two
  distinct domains ($v_c=2$) that vertices $c$, $1$, $2$ and $3$ belong to. In (b) $v_c=3$ if vertices
  $1$ and $2$ belong to the same domain; otherwise $v_c=3$.}
\end{figure}

\section{Efficient Metropolis update and numerical simulations}
\label{sec:numerics} We used a Monte Carlo method to sample typical
random graphs resulting from the POVM and compute their properties.
All Archimedean lattices can be embedded on rectilinear grids,
enabling simple labeling of sites and their
connections~\cite{SudingZiff}. The simulations utilize a generalized
Hoshen-Kopelman algorithm~\cite{HoshenKopelman} to identify
domains~\cite{WeiAffleckRaussendorf11,WeiAffleckRaussendorf12}. {Due
to the entanglement in the AKLT states, the local POVM outcomes are
correlated which is fully taken into account in our simulations. In
particular, to sample typical POVM outcomes ${\cal A}$ correctly, we
use a Metropolis method to update configurations. For each site we
attempt to flip the {\it type} (either $x$, $y$ or $z$) to one of
the other two equally and accept the flip with a probability
\begin{equation}
\label{eqn:pacc} p_{\rm acc}=\min\big\{1, 2^{|V'|-|{\cal
E}'|-|V|+|{\cal E}|}\big\},
\end{equation}
where $|V|$ and $|{\cal E}|$ denote the number of domains and
inter-domain or external edges (without the edge modulo-2
operation), respectively, before the flip, and similarly $|V'|$ and
$|{\cal E}'|$ for the flipped
configuration~\cite{WeiAffleckRaussendorf12}. We note that in terms
of the set of {\it internal\/} edges, denoted by ${\cal E}_I$, we
have that $|{\cal E}|+|{\cal E}_I|={\rm const.}$ Then the above
equation can be replaced by $p_{\rm acc}=\min\big\{1, 2^{|V'|+|{\cal
E}_I'|-|V|-|{\cal E}_I|}\big\}$.

The probability in Eq.~(\ref{eqn:pacc}) involves global quantities
$|{\cal E}|$ and $|V|$. But do we have to compute $|{\cal E}|$,
$|V|$, $|{\cal E}'|$ and $|V'|$ every time when we attempt to flip
the type of a site, as was done in
Refs.~\cite{WeiAffleckRaussendorf11,WeiAffleckRaussendorf11,DarmawanBrennenBartlett}?
The probability ratio can actually be replaced by a quasi-local
quantity. Denote by $e_I$ the number of internal edges impinging on
a vertex $c$ and $v_c$ the total number of distinct domains that the
site $c$ and its neighbors belong to. The quantity $e_I$ can be
straightforwardly checked locally by comparing the POVM type on $c$
and those on its neighbors; see Fig.~\ref{fig:wolff}. To count the
distinct domains, we first apply a Wolff algorithm~\cite{Wolff} to
grow domains on $c$ and the remaining sites of its neighbors (that
are not yet included in a domain) and then check the total number of
distinct domains. This can be done quasi-locally and in the worst
case scenario the largest domain has a size logarithmic in the
system size. Similarly, we can compute the $e_I'$ and $v_c'$ when
the POVM type on $c$ is flipped. Thus, we arrive at
\begin{equation}
\label{eqn:pacc3} p_{\rm acc}=\min\big\{1, 2^{v_c'+e_I'-v_c-e_I}
\big\}.
\end{equation}
With such a local update, the system size can reach to $N=L^2$ with
$L\sim1000$ without any difficulty. (Results of $L\sim500$ and $\sim
1000$ are illustrated, e.g., in Figs.~\ref{fig:statSO}
and~\ref{fig:statCross}, and are consistent with those from smaller
sizes.)

In the square-octagon and cross lattices, all the $3^N$ possible
POVM outcome configurations can occur. But in the case of the star
lattice, due to geometric frustration, the three sites in a triangle
cannot share the same POVM outcome. This is taken care of by
excluding the initial random assignment to be such configurations
and by forbidding transition to such configurations during the
simulations.

\begin{figure}
\vspace{-0.5cm} {\includegraphics[width=8.5cm]{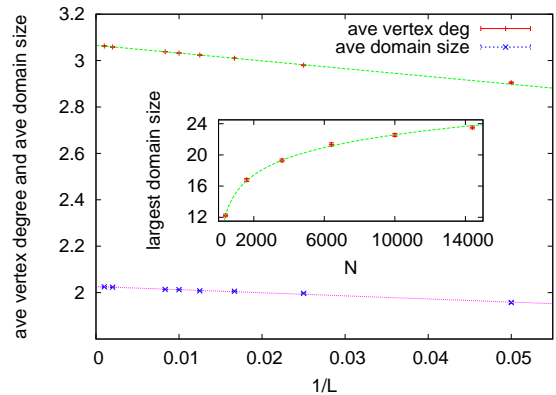}}
\vspace{-1cm}
 \caption{(color online)  Average domain size,
 average degree of a vertex, and the largest domain size (inset) in the typical graphs
 vs. $L=20,40,60,80,100,120,500,1000$, with $N= L^2$ being the total number
 of sites for the square-octagon lattice.  Inset: The largest domain size scales with $N$
 logarithmically.} \label{fig:statSO}
\end{figure}

\begin{figure}
 \begin{tabular}{l}
    (a)
\vspace{-1cm}\\ {\includegraphics[width=8.5cm]{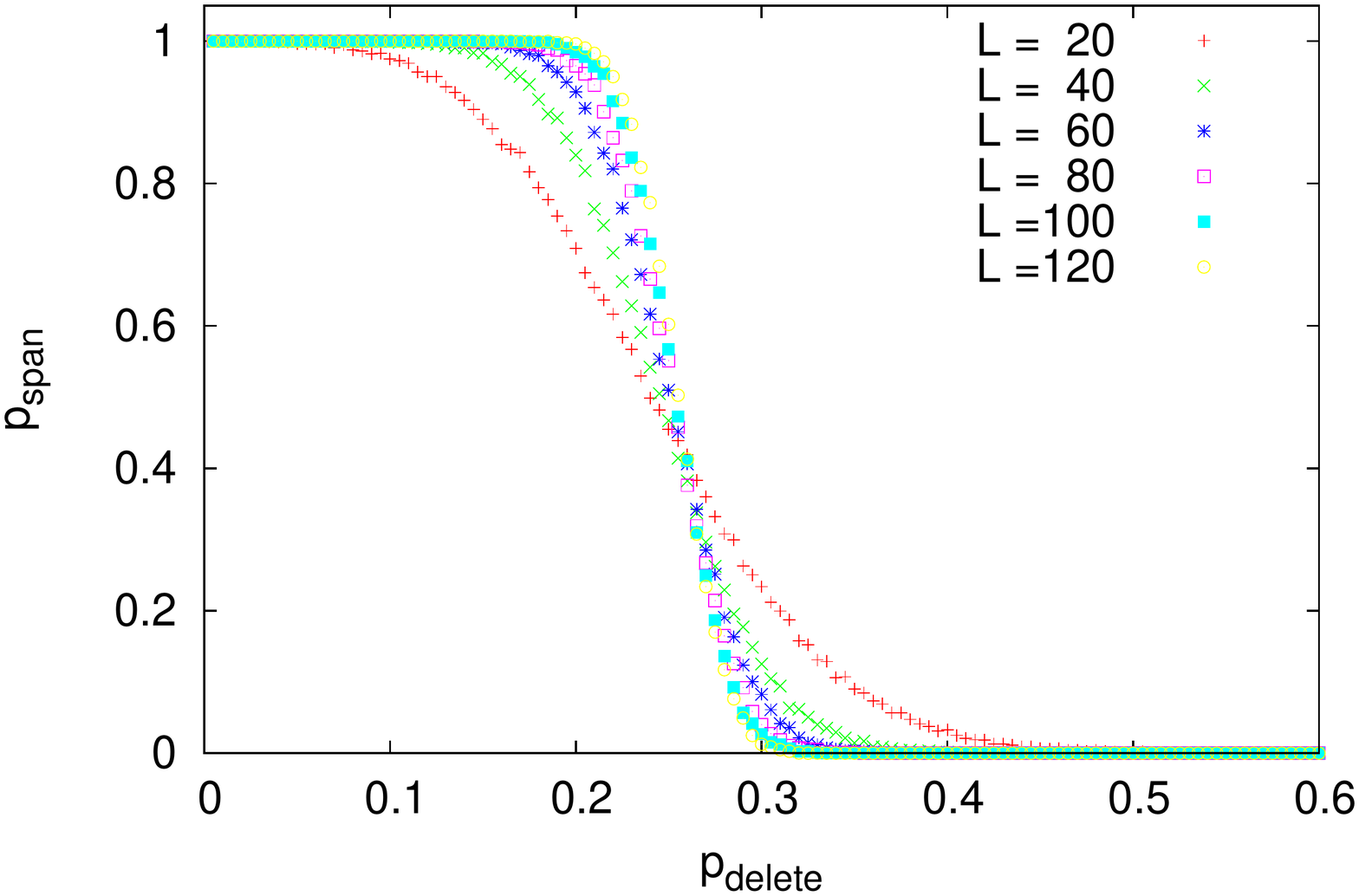}}
 \\
 (b)
 \vspace{-1cm}\\ {\includegraphics[width=8.5cm]{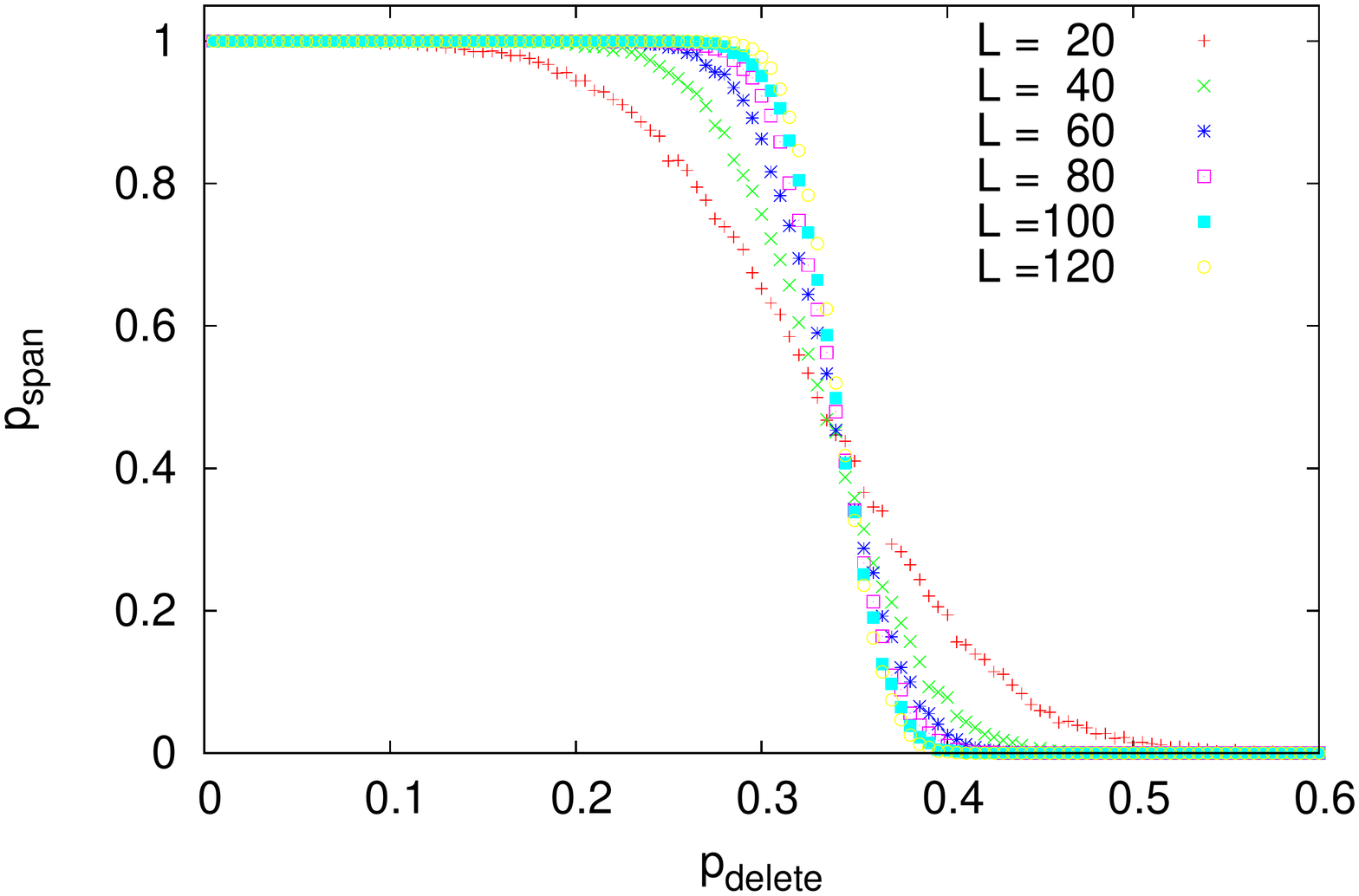}}
  \end{tabular}
        \caption{(color online)  Percolation study of the random graphs of domains resulting from the
         original { square-octagon} lattice: probability of
        a spanning cluster $p_{\rm span}$ vs. that to delete an edge  $p_{\rm
        delete}$ with $N=L^2$.  (a) Site percolation and (b) bond percolation.
         } \label{fig:percoSO}
\end{figure}

\begin{figure}
\vspace{-0.5cm} {\includegraphics[width=8.5cm]{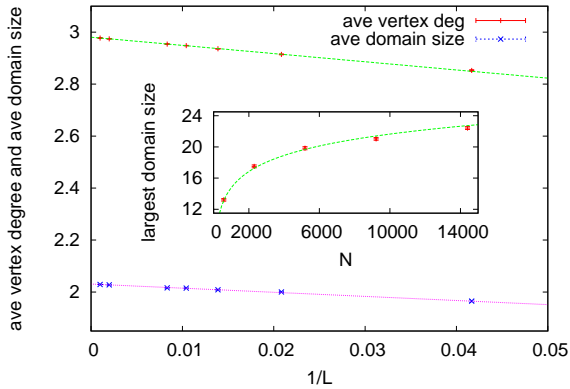}}
\vspace{-1cm}
 \caption{(color online)  Average domain size,
 average degree of a vertex, and the largest domain size (inset) in the typical graphs
 vs. $L=24,48,72,96,120,504,1008$, with $N= L^2$ being the total number
 of sites for the cross lattice.  Inset: The largest domain size scales with $N$
 logarithmically.} \label{fig:statCross}
\end{figure}

\begin{figure}
 \begin{tabular}{l}
    (a)
\vspace{-1cm} \\{\includegraphics[width=8.5cm]{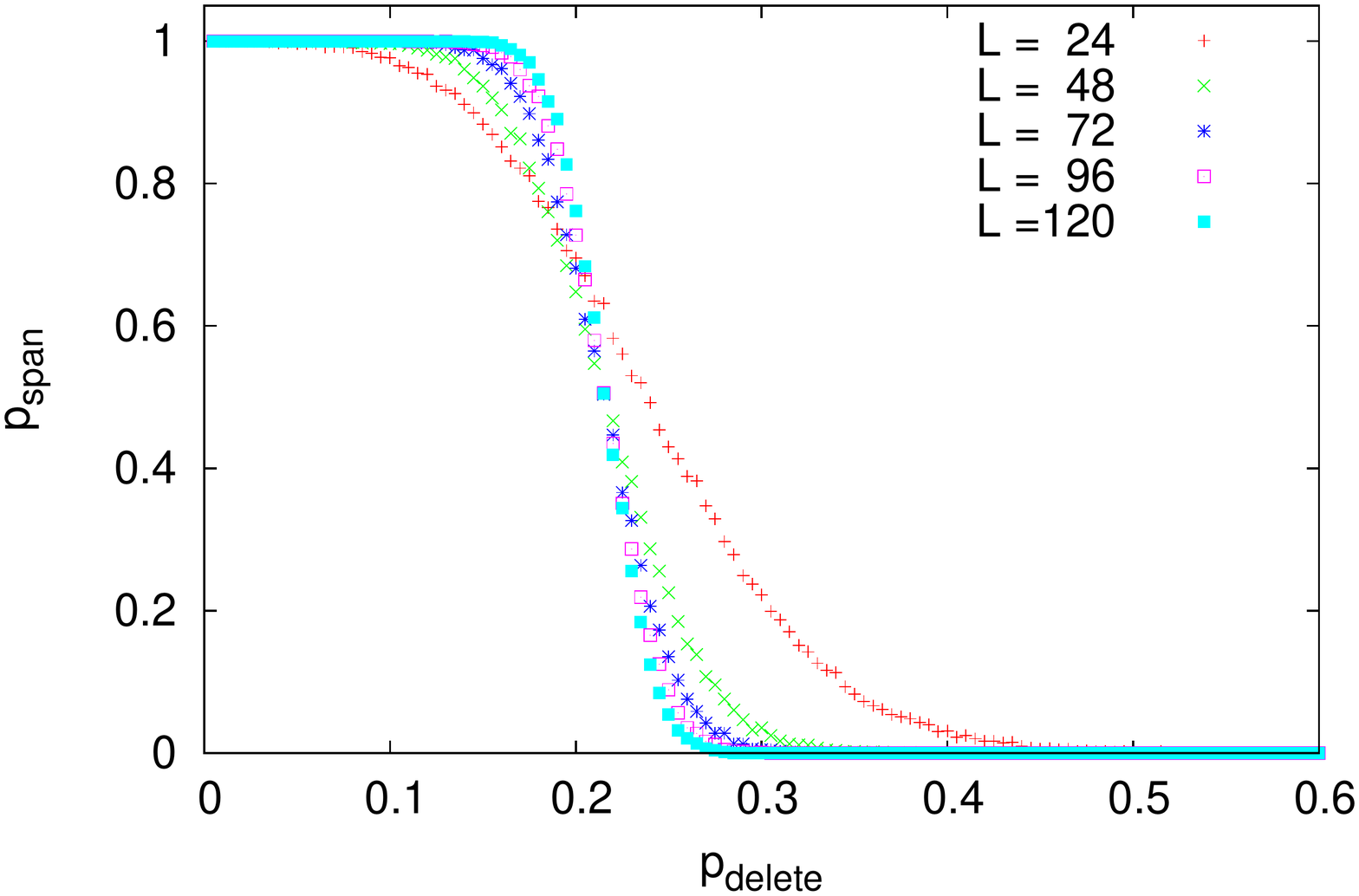}}
 \\
 (b)
 \vspace{-1cm}\\ {\includegraphics[width=8.5cm]{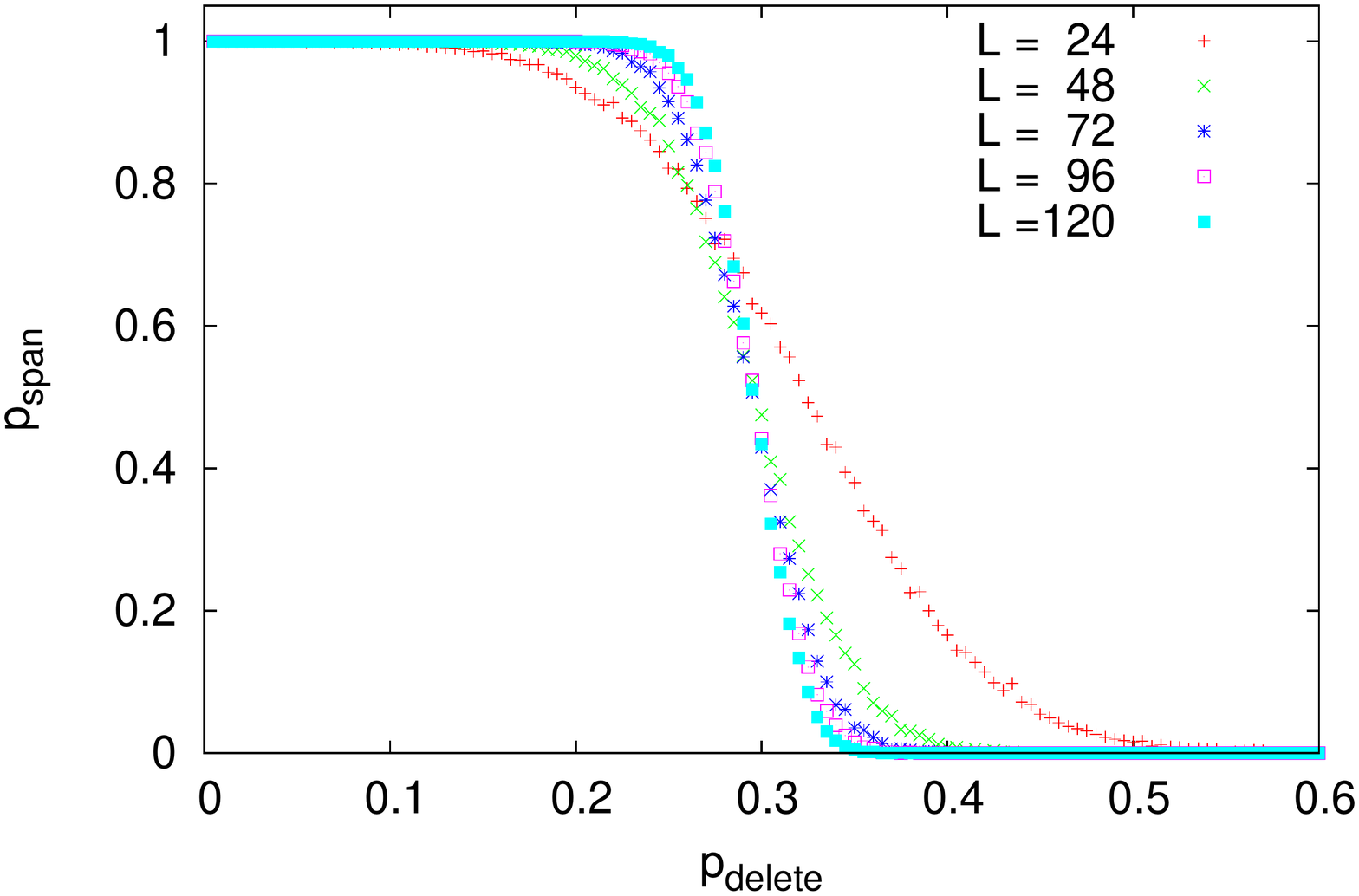}}
  \end{tabular}
        \caption{(color online)  Percolation study of the random graphs of domains resulting from the
         original { cross} lattice: probability of
        a spanning cluster $p_{\rm span}$ vs. that to delete an edge  $p_{\rm
        delete}$ with $N=L^2$.  (a) Site percolation and (b) bond percolation.
         } \label{fig:percoCross}
\end{figure}

\subsection{Square-octagon and cross lattices}
For the square-octagon and cross lattices, we have tested
that all the typical graphs of the post-POVM graph states are
percolated and deep in the supercritical phase, by randomly deleting
 a fraction of vertices or edges from  $G({\cal{A}})$.  For the square-octagon lattice,
the average domain size is $2.025(2)$ the average degree
$\bar{d}\approx3.065(1)$ (see Fig.~\ref{fig:statSO}). There remain
macroscopic number of vertices and edges: the number of vertices
$|\bar{V}|\approx 0.488(1) N$, the number of edges
$|\bar{E}|\approx0.748(1) N$. The typical graphs after the POVM
reside deep in the supercritical phase of percolation, as evidenced
by the fact that the threshold for site deletion is $p^{\rm
[site]*}_{\rm delete} \approx 0.26(1)$ and the threshold for edge
deletion is $p^{\rm [bond]*}_{\rm delete} \approx 0.35(1)$; see
Fig.~\ref{fig:percoSO}. The corresponding site and bond percolation
thresholds  ($p_{\rm th}=1-p_{\rm delete}$) are $0.74(1)$ and
$0.65(1)$, respectively.

For the cross lattice, the average domain size is $2.029(2)$, the
average degree $\bar{d}\approx 2.980(1)$ (see
Fig.~\ref{fig:statCross}). Similar to the square-octagon case, there
also remain macroscopic number of vertices and edges: the number of
vertices $|\bar{V}|\approx 0.485(1) N$, the number of edges
$|\bar{E}|\approx0.723(1) N$. Furthermore, the typical graphs after
the POVM reside deep in the supercritical phase of percolation, as
evidenced by the fact that  the threshold for site deletion is
$p^{\rm [site]*}_{\rm delete} \approx 0.21(1)$ and the threshold for
edge deletion is $p^{\rm [bond]*}_{\rm delete} \approx 0.29(1)$; see
Fig.~\ref{fig:percoCross}.  The corresponding site and bond
percolation thresholds  ($p_{\rm th}=1-p_{\rm delete}$) are
$0.79(1)$ and $0.71(1)$, respectively. Table~\ref{tbl:graph}
contains a summary of all these properties, including those from the
honeycomb AKLT state.

Let us examine the consequence of the above results in the context
of quantum computational universality. Whether or not typical graph
states $|G({\cal{A}})\rangle$ are universal resources hinges solely
on connectivity properties of $G({\cal{A}})$ (provided the graph is
macroscopic), and is thus a percolation problem~\cite{Perc}. Two
simple criteria can be used to determine whether the AKLT states on
the trivalent lattices ${\cal L}$ are
universal~\cite{WeiAffleckRaussendorf11,WeiAffleckRaussendorf12}:
\begin{enumerate}
  \item[C1]{\label{C1}The distribution of the number of sites in a domain (i.e. domain size)
   is {\em{microscopic}}, i.e.,  the largest domain size can at most scale logarithmically with
     the total number of sites $|V({\cal{L}})|$ in the large ${\cal{L}}$ limit.}
  \item[C2]{\label{C2}The probability of the existence of a path through $G({\cal{A}})$ from the left to the right
  (or top to bottom) approaches unity in the limit of large ${\cal{L}}$.}
\end{enumerate}
 Condition~C1 ensures that the
graph $G({\cal{A}})$ remains macroscopic, i.e., possesses
macroscopic number of vertices and edges, if the original
${\cal{L}}$ was.  Condition~C2 ensures that the system is in the
supercritical phase with a macroscopic spanning cluster of domains,
and hence
 there exist sufficient number of paths that can be used to
simulate evolution of qubits (such as single qubit gates) and their
interactions (such as the CNOT gate). For a large initial lattice
${\cal{L}}$ the random graph state $|G({\cal{A}})\rangle$ resulting
from the POVM can thus be efficiently reduced to a large
two-dimensional cluster state~\cite{WeiAffleckRaussendorf12}.

It is evident that the results of our numerical simulations
demonstrate that both conditions C1 and C2 are satisfied in the
cases of the square-octagon and cross lattices. This shows that they
are both universal resources for MBQC.

\begin{figure*}
 \begin{tabular}{l}
    (a)\\
  \includegraphics[width=0.7\textwidth]{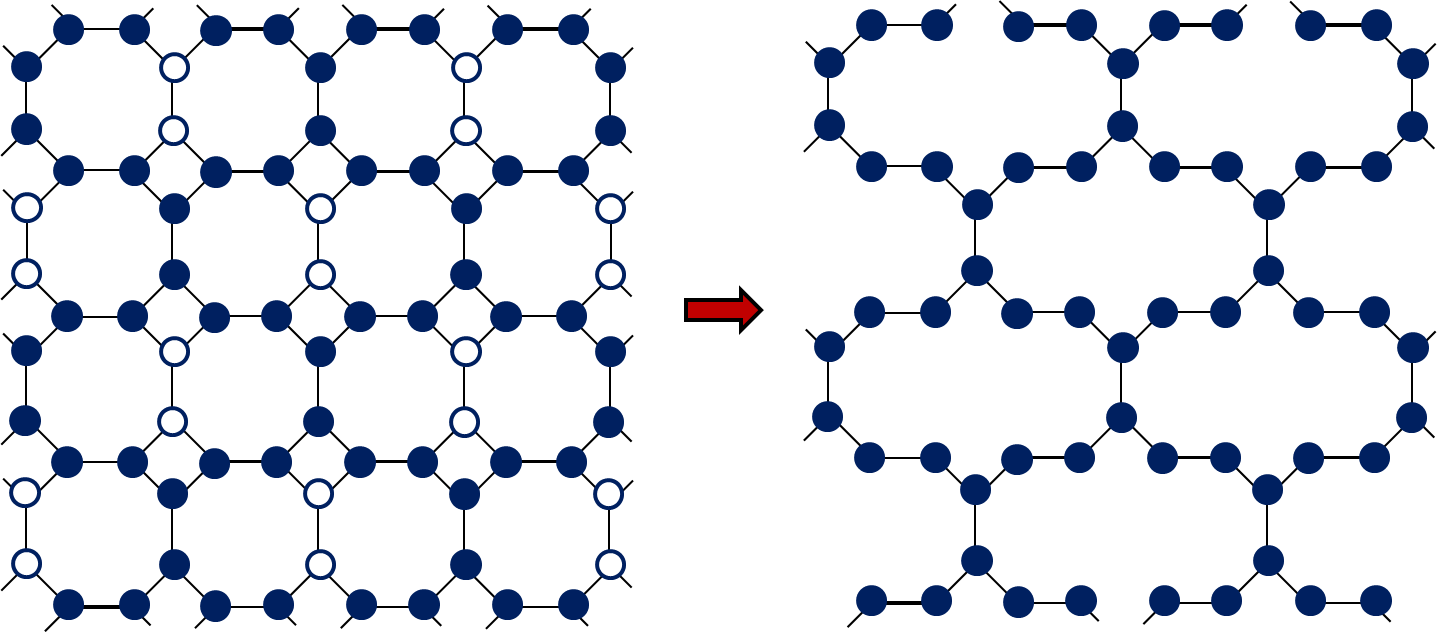}
   \vspace{0.5cm}\\
   (b)\\
   \includegraphics[width=0.7\textwidth]{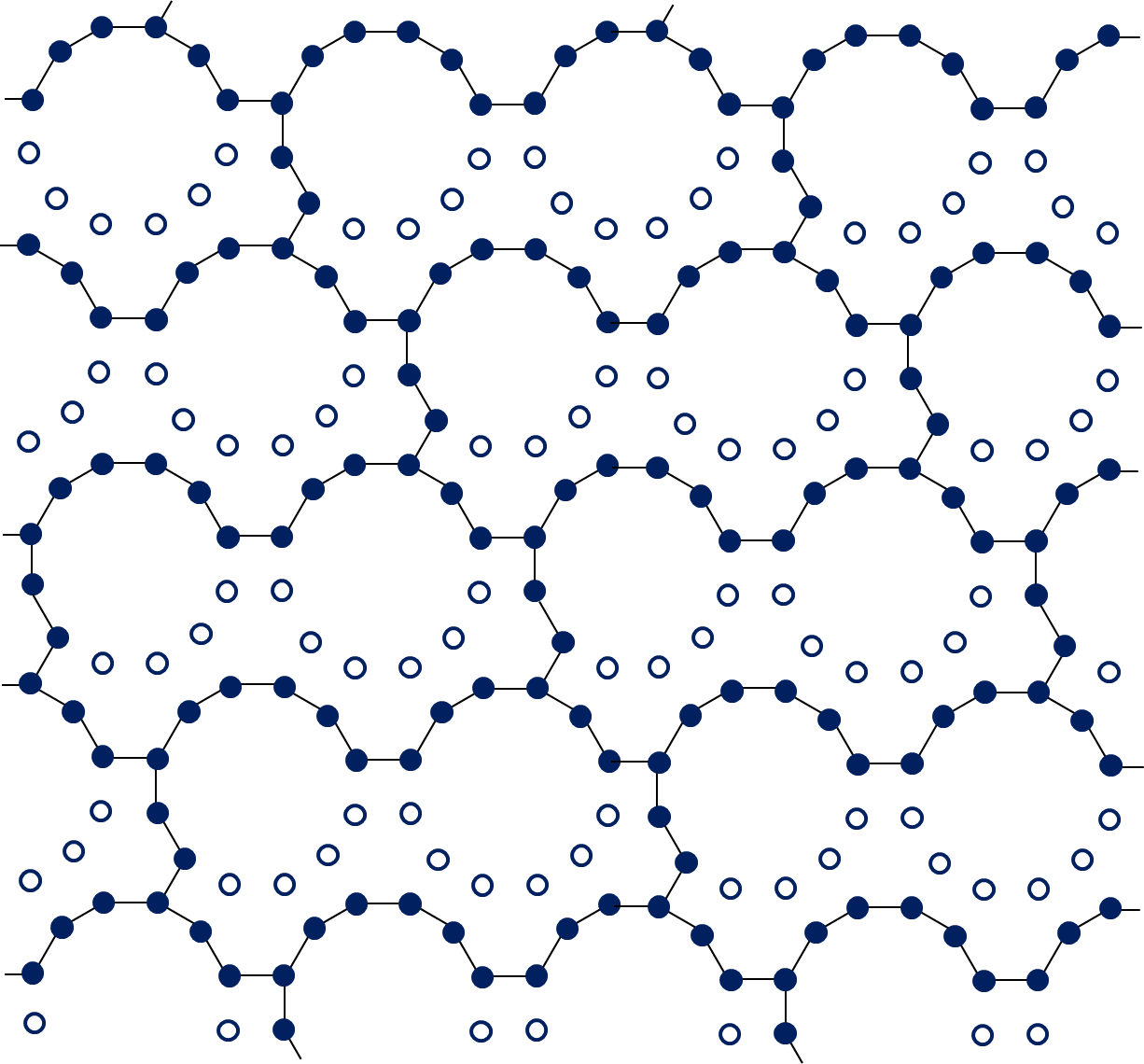}
   \end{tabular}
  \caption{\label{fig:socross}(color online) Illustration of the key local measurement
  in converting the cluster states on (a) the square-octagon  and (b) the cross lattices to a
  cluster state on the honeycomb lattice. The hollow circles indicate the locations where Pauli Z measurement
  is performed. The picture on the r.h.s. in (a) shows the resulting graph after removing the Z measured sites, which
  is topologically equivalent to a honeycomb lattice, except with additional sites of vertex degree 2.
  In (b) the resultant graph is shown by removing the edges due to Z measurements on the sites indicated by hollow circles.
  In both cases, Pauli X measurements can be applied to
  those sites of vertex degree 2 to further reduce to the honeycomb lattice. }
\end{figure*}
\subsection{The star lattice}
For the star lattice, however, none of the typical POVM
configurations that we generated (for $L\ge 48$) yield percolated
graphs, i.e., they reside in the subcritical phase, even without any
deletion of vertices or edges. This shows that there does not exist
a computational backbone (formed by domains) that extends to a
two-dimensional grid and thus there are not sufficient paths to
simulate evolution of qubits and their interaction. Hence, the
resultant graph states cannot be used to perform universal quantum
computation. We should stress that for a finite system the POVM can,
albeit with a probability exponentially small with the system size,
yield graph states (e.g. a cluster state on the original star
lattice) whose graphes are indeed percolated, consistent with our
simulations. However, we are concerned with the ability to generate
universal resource states in the large system limit. In this limit,
the probability that one obtains graph states that are universal
approaches zero, consistent with our numerical simulations as we
increases the system size.

\begin{figure*}
  \includegraphics[width=0.9\textwidth]{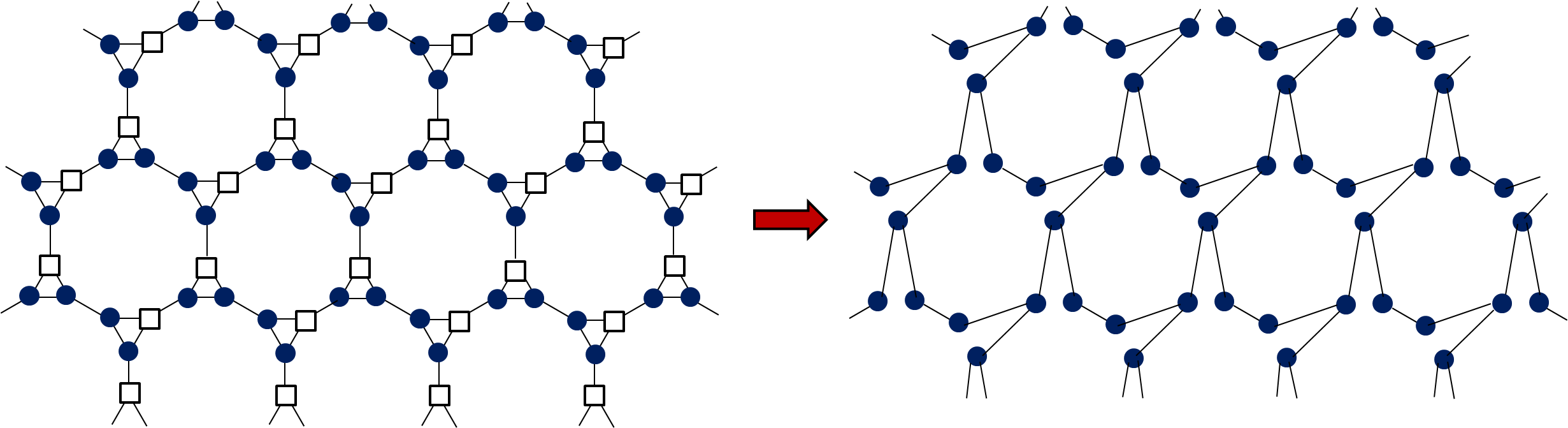}
  \caption{\label{fig:star_cluster}(color online) Illustration of the key local measurement
  in converting the cluster state on the star lattice to a
  cluster state on the honeycomb lattice. The hollow squares indicate the locations where Pauli Y measurement
  is performed. The action of Pauli Y measurement is to perform local complementation followed by removal of the vertex measured.
  The resultant graph is shown on the r.h.s., which is topologically equivalent to a honeycomb lattice. Pauli X measurements can be applied to
  those sites of vertex degree 2 to further reduce to the honeycomb lattice. }
\end{figure*}
This lack of universality for the resultant typical graph states is
due to the antiferromagnetic property of the original AKLT state and
the geometric frustration in the star lattice. An intuitive
understanding is as follows. Consider a triangle and because of
geometric frustration, the POVM outcome with $(x,x,x)$, $(y,y,y)$
and $(z,z,z)$ cannot occur. There are 6 outcomes with all three
labels be different  (case 1) and 18 outcomes with only two labels
being the same and the third being distinct (case 2). In the former
case the three edges remain, whereas in the latter case two of the
three edges are removed (and the remaining one becomes an internal
edge). This gives rise to an average probability of $p_{\rm
delete}^{\rm [bond]}= 1/4*0 +3/4* 2/3= 1/2$ to remove an edge in an
triangle. The edges that connect neighboring triangles either become
internal edges or remain external edges (see Fig.~\ref{fig:star}).
Hence for the purpose of percolation, we can consider contracting
these edges and thus the star lattice becomes a Kagome lattice,
which has a bond percolation threshold $p_{\rm th}^{\rm
[bond]}\approx 0.5244$. Thus deleting an edge with probability $1/2$
will make the graph become un-percolated. We note that consideration
of correlated removal of edges should be taken into account. Our
numerical simulations do take into account all correlations in the
configuration (and hence correlations in removing edges) and confirm
that the graphs resulting from the POVM are in the subcritical
phase. This implies that the AKLT state on the star lattice under
the POVM cannot produce a graph state that is universal for MBQC,
which fails to satisfies Condition C2 (percolation). In contrast,
the N\'eel phase in the deformed AKLT model on the honeycomb
lattice~\cite{NiggemannKlumperZittarz} was argued not universal due
to the breaking of Condition C1 (non-macroscopic
domains)~\cite{DarmawanBrennenBartlett}. We should emphasize that we
have not ruled out the possibility that other POVM's might  enable
the star-lattice AKLT state for universal MBQC, even though we
believe that it is unlikely. On another note, frustrated lattices
such as the star and Kagome lattices are of particular interest in
condensed matter physics, as they can host exotic phases of matter,
such as the spin
liquids~\cite{YaoKivelson,YangParamekantiKim,YanHuseWhite}. Here, we
have witnessed evidence that such frustration may prevent a certain
type of universal resource states to exist.

 \section{Concluding remarks}
 \label{sec:conclude}
 Given that the spin-3/2 AKLT state on
the honeycomb lattice has been
 shown to be a universal resource for MBQC, a natural question arises whether other AKLT states can also
 be universal.
 We have thus investigated MBQC with the spin-3/2 AKLT states
on all four trivalent Archimedean lattices and found that they are
also universal on two other lattices: the square-octagon $(4,8^2)$
and the cross  $(4,6,12)$. Furthermore, the same POVM performed on
the AKLT state on the star lattice $(3,12^2)$ yields typical graph
states whose graphs are in subcritical phase (in the sense of
percolation), and this suggests the AKLT state  on the star lattice
may not be universal for MBQC. However, it might be possible that
other POVM's could turn this AKLT state into a universal resource,
which, remains an open question.

\medskip \noindent {\bf Acknowledgment.} The author thanks Ian Affleck and Robert Raussendorf for
useful discussions. This work was supported by the National Science
Foundation under Grants No. PHY 1314748 and No. PHY 1333903.

\appendix
\section{Cluster states on trivalent Archimedean lattices}
The cluster state on one of the four trivalent Archimedean lattices,
namely, the honeycomb lattice, was shown by Van den Nest et al. to
be universal for MBQC~\cite{Universal}. Their approach is to use
local Pauli measurements and their effects on the original graph is
to transform it with certain rules~\cite{Universal,Hein2}. For
example, a Pauli Z measurement on a qubit corresponds to deleting
edges incident to the vertex that the qubit resides on, hence
disentangling the qubit (whose vertex is removed as well). A Pauli Y
measurement on a qubit corresponds to first implementing the local
complementation on the neighboring vertices, followed by removing
the measured vertex and edges incident on it. Using these rules, we
can prove that the all the cluster states on the square-octagon,
cross and star lattices can be converted to a cluster state on the
honeycomb lattice, which is universal. The key step for all three
cases are illustrated in
Figs.~\ref{fig:socross}\&~\ref{fig:star_cluster}.


\begin{thebibliography}{99}
\bibitem{NielsenChuang00}
M. Nielsen and I. Chuang, {\sl Quantum Computation and Quantum Information\/}
(Cambridge Univ. Press, 2000).
\bibitem{Oneway}
R. Raussendorf and H. J. Briegel, Phys. Rev. Lett. {\bf 86}, 5188
(2001).

\bibitem{Oneway2}
H. J. Briegel, D. E. Browne, W. D\"ur, R. Raussendorf, and M. Van
den Nest, Nature Phys. {\bf 5}, 19 (2009).


\bibitem{RaussendorfWei12}
R. Raussendorf and T.-C. Wei,
Annu. Rev. Condens. Matter Phys. {\bf 3}, pp.239-261 (2012).

\bibitem{VandenNestDurVidalBriegel07}
M. Van den Nest, W. D\"ur, G. Vidal, and H. J. Briegel,
Phys. Rev. A {\bf 75}, 012337 (2007).

\bibitem{VandenNest}
M. Van den Nest, W. D\"ur, A. Miyake, and H. J. Briegel,
New J. Phys. {\bf 9}, 204 (2007).

\bibitem{Gross1}
D. Gross, S.T. Flammia, and J. Eisert, Phys. Rev. Lett.
\textbf{102}, 190501 (2009); M.J. Bremner, C. Mora, and A. Winter,
{\it ibid\/} {\bf 102}, 190502 (2009).

\bibitem{Cluster}
H. J. Briegel and R. Raussendorf,  Phys. Rev. Lett. {\bf 86}, 910 (2001).

\bibitem{Nielsen}
M. A. Nielsen,  Rep. Math. Phys. {\bf 57}, 147 (2006).




\bibitem{Gross}
D. Gross and J. Eisert,  Phys. Rev. Lett. {\bf 98}, 220503 (2007).


\bibitem{Verstraete}
F. Verstraete and J. I. Cirac, Phys. Rev. A {\bf 70}, 060302(R)
(2004).

 \bibitem{Chen}
X. Chen, B. Zeng, Z.-C. Gu, B. Yoshida, and I. L. Chuang, Phys. Rev.
Lett. {\bf 102}, 220501 (2009).





\bibitem{AKLT}
I. Affleck, T. Kennedy, E. H. Lieb, and H. Tasaki,  Phys. Rev. Lett.
{\bf 59}, 799 (1987).
\bibitem{AKLT2} I. Affleck, T. Kennedy, E. H. Lieb, and H. Tasaki,
Comm. Math. Phys. {\bf 115}, 477 (1988).

\bibitem{Brennen}
G. K. Brennen and A. Miyake,  Phys. Rev. Lett. {\bf 101}, 010502 (2008).

\bibitem{Haldane}
F. D. M. Haldane,
Phys. Lett. A {\bf 93}, 464-468 (1983).

\bibitem{Haldane2}
F. D. M. Haldane,
Phys. Rev. Lett. {\bf
50}, 1153-1156 (1983).


\bibitem{KLT}
T. Kennedy, E. H. Lieb, and H. Tasaki, J. of Stat. Phys. {\bf 53},
383 (1988).

\bibitem{Resch}
R. Kaltenbaek , J. Lavoie,
 B. Zeng,
 S. D. Bartlett, and K. J. Resch, Nature Phys. {\bf 6}, 850 (2010).



\bibitem{Bartlett} S. D. Bartlett, G. K. Brennen, A. Miyake, and J. M. Renes,
 Phys. Rev. Lett.
{\bf 105}, 110502 (2010).

\bibitem{Miyake}
A. Miyake, Phys. Rev. Lett. {\bf 105}, 040501 (2010).

\bibitem{ElseSchwarzBartlettDoherty}
D. V. Else, I. Schwarz, S. D. Bartlett, and A. C. Doherty,
Phys. Rev. Lett. {\bf 108}, 240505 (2012).



\bibitem{WeiAffleckRaussendorf11}
 T.-C. Wei, I. Affleck, and R. Raussendorf,
  Phys. Rev. Lett. {\bf 106}, 070501 (2011).

\bibitem{Cai10} J.-M. Cai, A. Miyake, W.
D\"ur, and H. J. Briegel, Phys. Rev. A {\bf 82}, 052309 (2010).




\bibitem{WeiRaussendorfKwek11} T.-C. Wei, R. Raussendorf, and L. C.
Kwek,
Phys. Rev. A {\bf 84}, 042333 (2011).

\bibitem{Miyake11} A. Miyake,  Ann. Phys. (Leipzig) {\bf 326}, 1656
(2011).

\bibitem{Param}
 S. A. Parameswaran, S. L. Sondhi, and D. P. Arovas,
Phys. Rev. B {\bf 79}, 024408 (2009).

\bibitem{Universal}
M. Van den Nest, A. Miyake, W. D\"ur, and H. J. Briegel,  Phys. Rev. Lett.
{\bf 97}, 150504 (2006).




\bibitem{WeiAffleckRaussendorf12}
 T.-C. Wei, I. Affleck, and R. Raussendorf,
  Phys. Rev. A {\bf 86}, 032328 (2012).


\bibitem{DarmawanBrennenBartlett}
A. S. Darmawan, G. K. Brennen, and S. D. Bartlett,
New J. Physics, {\bf 14}, 013023 (2012).


\bibitem{Hein}
M. Hein, J. Eisert, and H.-J. Briegel,   Phys. Rev. A {\bf 69}, 062311 (2004).

\bibitem{Stabilizer}
D. Gottesman, {\it Stabilizer Codes and Quantum Error Correction\/}, Ph.D.
Thesis, Caltech (1997).

\bibitem{Browne}
D. E. Browne, M. B. Elliott, S. T. Flammia, S. T. Merkel, A. Miyake,
A. J. Short, New J. Physics {\bf 10}, 023010 (2008).

\bibitem{SudingZiff}
P. N. Suding and R. M. Ziff, Phys. Rev. E  {\bf 60}, 275 (1999).
\bibitem{HoshenKopelman}
J. Hoshen and R. Kopelman,  Phys. Rev. B {\bf 14}, 3438 (1976).

\bibitem{Wolff}
U. Wolff, Phys. Rev. Lett. {\bf 62}, 361 (1989).

\bibitem{Perc}
  R. Durrett, {\em{Random Graph Dynamics}}, Cambridge University Press (2007).



\bibitem{NiggemannKlumperZittarz}
H. Niggemann, A. Kl\"umper, and J. Zittartz,
Z. Phys. B {\bf 104}, 103 (1997).

\bibitem{YaoKivelson}
H. Yao and S. A. Kivelson, Phys. Rev. Lett. {\bf 99}, 247203 (2007).

\bibitem{YangParamekantiKim}
B.-J. Yang, A. Paramekanti, and Y. B. Kim, Phys. Rev. B {\bf 81},
134418 (2010).

\bibitem{YanHuseWhite}
S. Yan, D. A. Huse, and S. R. White, Science {\bf 332}, 1173 (2011).

\bibitem{Hein2}
M. Hein, W. D\"ur, J. Eisert, R. Raussendorf, M. Van den Nest, and
H.-J. Briegel, eprint quant-ph/0602096.
\end{thebibliography}
\end{document}